\documentstyle[12pt]{article}

\font\twlgot =eufm10 scaled \magstep1 \font\egtgot =eufm8
\font\sevgot =eufm7 \font\twlmsb =msbm10 scaled \magstep1
\font\egtmsb =msbm8 \font\sevmsb =msbm7

\newfam\gotfam
\def\pgot{\fam\gotfam\twlgot}
\textfont\gotfam\twlgot \scriptfont\gotfam\egtgot
\scriptscriptfont\gotfam\sevgot
\def\got{\protect\pgot}
\newfam\msbfam
\textfont\msbfam\twlmsb \scriptfont\msbfam\egtmsb
\scriptscriptfont\msbfam\sevmsb
\def\Bbb{\protect\pBbb}
\def\pBbb{\relax\ifmmode\expandafter\Bb\else\typeout{You cann't use
Bbb in text mode}\fi}
\def\Bb #1{{\fam\msbfam\relax#1}}

\newcommand{\gd}{{\got d}}

\def\thebibliography#1{\section*{References}\list
  {[\arabic{enumi}]}{\settowidth\labelwidth{#1}\leftmargin\labelwidth
    \advance\leftmargin\labelsep
    \usecounter{enumi}}
    \def\newblock{\hskip .11em plus .33em minus .07em}
    \sloppy\clubpenalty4000\widowpenalty4000
    \sfcode`\.=1000\relax}

\def\op#1{\mathop{\fam0 #1}\limits}

\newcommand{\id}{{\rm Id\,}}

\newcommand{\beq}{\begin{equation}}
\newcommand{\eeq}{\end{equation}}
\newcommand{\ben}{\begin{eqnarray}}
\newcommand{\een}{\end{eqnarray}}
\newcommand{\be}{\begin{eqnarray*}}
\newcommand{\ee}{\end{eqnarray*}}
\newcommand{\bea}{\begin{eqalph}}
\newcommand{\eea}{\end{eqalph}}
\newcommand{\cA}{{\cal A}}

\newcommand{\cP}{{\cal P}}

\newcommand{\cL}{{\cal L}}

\newcommand{\cE}{{\cal E}}

\newcommand{\cQ}{{\cal Q}}
\newcommand{\cF}{{\cal F}}
\newcommand{\cS}{{\cal S}}

\newcommand{\cO}{{\cal O}}

\newcommand{\cG}{{\cal G}}

\newcommand{\cN}{{\cal N}}
\newcommand{\bL}{{\bf L}}

\newcommand{\al}{\alpha}

\newcommand{\bt}{\beta}
\newcommand{\dl}{\delta}
\newcommand{\la}{\lambda}
\newcommand{\La}{\Lambda}
\newcommand{\f}{\phi}

\newcommand{\m}{\mu}

\newcommand{\g}{\gamma}

\newcommand{\th}{\theta}

\newcommand{\vt}{\vartheta}

\newcommand{\up}{\upsilon}
\newcommand{\lng}{\langle}
\newcommand{\rng}{\rangle}

\newcommand{\si}{\sigma}
\newcommand{\Si}{\Sigma}
\newcommand{\w}{\wedge}

\newcommand{\wh}{\widehat}
\newcommand{\ol}{\overline}
\newcommand{\dr}{\partial}
\newcommand{\ar}{\op\longrightarrow}

\newcommand{\ot}{\otimes}

\newcounter{eqalph}
\newcounter{equationa}
\newcounter{remark}
\newcounter{example}
\newcounter{theorem}
\newcounter{proposition}
\newcounter{lemma}
\newcounter{corollary}
\newcounter{definition}
\def\theremark{\arabic{remark}}

\def\thedefinition{\arabic{definition}}

\newenvironment{eqalph}{\stepcounter{equation}
\setcounter{equationa}{\value{equation}} \setcounter{equation}{0}

\begin{eqnarray}}{\end{eqnarray}
\setcounter{equation}{\value{equationa}}}

\newcommand{\mar}[1]{}

\begin{document}
\hbox{}

{\parindent=0pt

{\large\bf Green function identities in Euclidean quantum field theory}
\bigskip

{\sc G. Sardanashvily}\footnote{E-mail: gennadi.sardanashvily@unicam.it}
\medskip

\begin{small}

{\sl Department of Theoretical Physics, Moscow State University,
117234 Moscow, Russia}

\bigskip
\bigskip

{\bf Abstract}

Given a generic Lagrangian system of even and odd fields, we
show that any infinitesimal transformation of its classical Lagrangian
yields the identities which Euclidean Green functions of quantum
fields satisfy. 

\end{small}

 }

\section{Introduction}

We aim to show
that, obeying the first variational formula (\ref{g107}), any
infinitesimal transformation of a classical Lagrangian field system
yields the identities (\ref{z62'}) which Euclidean Green
functions of quantum fields satisfy.

A generic Lagrangian system of even and odd fields is
considered [1-3]. It is degenerate, and must be brought into a
nondegenerate Lagrangian system in order to be quantized in the
framework of a perturbative QFT. An Euler--Lagrange operator 
of a degenerate Lagrangian system satisfies nontrivial Noether identities.
They need not be independent, but obey the first-stage Noether identities,
which in turn are subject to the second-stage ones, and so on. Being
finitely generated, these Noether identities are
parameterized by the modules of antifields. The Noether's
second theorem states the relation between these Noether identities and
the reducible gauge supersymmetries of a degenerate Lagrangian
system parameterized by ghosts. In the framework of the
Batalin--Vilkovisky (BV) quantization of a degenerate Lagrangian
system [4-7], its original Lagrangian 
is extended to the above mentioned ghosts and antifields in order to
satisfy the so-called classical master equation. Replacing antifields
with gauge fixed terms, one comes to a nondegenerate gauge-fixing
Lagrangian which is quantized. Instead
of a gauge symmetry of an original Lagrangian, this Lagrangian possesses
the variational BRST supersymmetry. 

Bearing in mind quantization, we consider a Lagrangian
field system on $X=\Bbb R^n$, $n\geq 2$, coordinated by $(x^\la)$.
It is described in algebraic terms of a graded commutative
$C^\infty(X)$-algebra
$\cP^0$ with generating elements
\mar{bb3}\beq
\{s^a, s^a_\la, s^a_{\la_1\la_2},\ldots, s^a_{\la_1\ldots\la_k},\ldots\}, 
\label{bb3}
\eeq
and the bigraded differential algebra  
$\cP^*$ of differential forms (the Chevalley--Eilenberg
differential calculus) over $\cP^0$ as an $\Bbb R$-algebra [1-3]. One can
think of generating elements (\ref{bb3}) of $\cP^0$ 
as being {\it sui generis} coordinates of even and odd fields and their
partial derivatives (jets) (Section 2). The symbol
$[a]=[s^a]=[s^a_{\la_1\ldots\la_k}]$ stands for their Grassmann parity.
In fact, $\cP^0$
is the algebra of polynomials in the graded elements (\ref{bb3})
whose coefficients are real smooth functions on $X$. The graded
commutative $\Bbb R$-algebra $\cP^0$ is provided with the even graded
derivations (called total derivatives) 
\mar{bb2}\beq
d_\la =\dr_\la + \op\sum_{0\leq|\La|} s_{\la+\La}^a\dr^\La_a, \qquad
d_\La=d_{\la_1}\cdots d_{\la_k}, \label{bb2}
\eeq
where $\La=(\la_1...\la_k)$, $|\La|=k$, and $\la+\La=
(\la,\la_1,\ldots,\la_k)$ are symmetric multi-indices.
One can think of even elements
\mar{0709}\beq
L=\cL(x^\la,s^a_\La) d^nx,
\qquad \dl L= ds^a\w \cE_a d^nx=\op\sum_{0\leq|\La|}
 (-1)^{|\La|} ds^a\w d_\La (\dr^\La_a L)d^nx
\label{0709}
\eeq
of the differential algebra $\cP^*$ as being a graded Lagrangian and its
Euler--Lagrange operator, respectively. By infinitesimal transformations
of a Lagrangian system are meant odd vertical contact graded derivations 
\mar{0672}\beq
\vt=\up^a\dr_a + \op\sum_{0<|\La|} d_\La\up^a\dr_a^\La
\label{0672}
\eeq
of the $\Bbb R$-algebra $\cP^0$. The Lie derivative
$\bL_\vt L$ of a Lagrangian $L$ (\ref{0709}) along such a derivation
obeys the first variational formula
\mar{g107}\beq
\bL_\vt L= \up^a\cE_a d^nx +d_\la J^\la d^nx, \label{g107}
\eeq
where $J^\la$ is a generalized Noether current. One says that $\vt$
(\ref{0672}) is a variational supersymmetry of a  Lagrangian $L$ if the
Lie derivative (\ref{g107}) is a divergence $\bL_\vt L=d_\la\si^\la
d^nx$, e.g., it vanishes.

From now on, let $(\cP^*,L)$ be either an original nondegenerate
Lagrangian system or the nondegenerate Lagrangian system brought from
an original degenerate one by means of the above mentioned
BV procedure. Let us quantize this Lagrangian system in
the framework of perturbative Euclidean QFT (Section 3). We suppose
that $L$ is a Lagrangian of Euclidean fields on $X=\Bbb R^n$. The key
point is that the algebra of Euclidean quantum fields $B_\Phi$ as like as
$\cP^0$ is graded commutative. It is generated by elements $\f^a_{x\La}$,
$x\in X$.  For
any $x\in X$, there is a homomorphism
\mar{z45}\beq
\g_x: f_{a_1\ldots a_r}^{\La_1\ldots\La_r} s^{a_1}_{\La_1}\cdots
s^{a_r}_{\La_r} \mapsto f_{a_1\ldots a_r}^{\La_1\ldots\La_r}(x)
\f^{a_1}_{x\La_1}\cdots \f_{x\La_r}^{a_r}, \qquad f_{a_1\ldots
a_r}^{\La_1\ldots\La_r}\in C^\infty(X), \label{z45}
\eeq
of the algebra $\cP^0$ of classical fields to the algebra $B_\Phi$
which sends the basis elements $s^a_\La\in \cP^0$ to the
elements $\f^a_{x\La}\in B_\Phi$, and replaces
coefficient functions $f$ of elements of $\cP^0$ with their
values $f(x)$ at a point $x$. Then a state $\lng.\rng$
of $B_\Phi$ is given by symbolic functional integrals
\mar{z49}\ben
&& \lng\f^{a_1}_{x_1}\cdots \f^{a_k}_{x_k}\rng=\frac{1}{\cN}\int
\f^{a_1}_{x_1}\cdots \f^{a_k}_{x_k} \exp\{-\int
\cL(\f^a_{x\La})d^nx\}\op\prod_x [d\phi_x^a], \label{z49}\\
&& \cN=\int \exp\{-\int \cL(\f^a_{x\La})d^nx\}\op\prod_x
[d\phi_x^a], \nonumber\\
&&
\cL(\f^a_{x\La})=\cL(x,\g_x(s^a_\La)), \nonumber
\een
which restart complete Euclidean Green functions in
the Feynman diagram technique. 

Due to homomorphisms (\ref{z45}), any graded derivation
$\vt$ (\ref{0672})
of $\cP^0$ induces the graded derivation
\mar{z50'}\beq
\wh \vt: \f^a_{x\La}\to (x,
s^a_\La)\to u^a_\La(x,s^b_\Si)\to 
u^a_\La(x,\g_x(s^b_\Si))=\wh \vt^a_{x\La}(\f^b_{x\Si}) \label{z50'}
\eeq
of the algebra of quantum fields $B_\Phi$ (Section 4).
With an odd parameter $\al$, let us consider the
automorphism
\mar{bb20}\beq
\wh U=\exp\{\al \wh \vt\}=\id +\al\wh \vt \label{bb20}
\eeq
of the algebra $B_\Phi$. This automorphism yields a new state
$\lng.\rng'$ of $B_\Phi$ given by the equalities
\mar{bb10}\ben
&& \lng \f^{a_1}_{x_1}\cdots \f^{a_k}_{x_k}\rng= \lng \wh
U(\f^{a_1}_{x_1})\cdots \wh U(\f^{a_k}_{x_k})\rng'=  \label{bb10}\\
&& \qquad \frac{1}{\cN'}\int \wh U(\f^{a_1}_{x_1})\cdots \wh
U(\f^{a_k}_{x_k})
\exp\{-\int \cL(\wh U(\f^a_{x\La}))d^nx\}\op\prod_x [d\wh
U(\phi_x^a)], \nonumber \\
&& \cN'=\int \exp\{-\int \cL(\wh U(\f^a_{x\La}))d^nx\}\op\prod_x
[d\wh U(\phi_x^a)]. \nonumber
\een
It follows from the first variational formula (\ref{g107}) that
\be
\int \cL(\wh U(\f^a_{x\La}))d^nx =\int
(\cL(\f^a_{x\La}) + \al \wh \vt_x^a\cE_{xa})d^nx,
\ee
where
$\cE_{xa}=\g_x(\cE_a)$ are the variational derivatives.
It is a property of symbolic functional integrals that
\mar{bb21}\beq
\op\prod_x[d\wh U(\phi_x^a)]=(1+\al\int \frac{\dr \wh
\vt^a_x}{\dr \f^a_x}d^nx)\op\prod_x[d\phi_x^a]=(1+\al {\rm Sp}(\wh
\vt)) \op\prod_x[d\phi_x^a]. \label{bb21}
\eeq
 Then the
equalities (\ref{bb10}) result in the identities 
\mar{z62'}\ben
&& \lng\wh \vt(\f^{a_1}_{x_1}\cdots \f^{a_k}_{x_k})\rng +
\lng\f^{a_1}_{x_1}\cdots \f^{a_k}_{x_k}({\rm Sp}(\wh \vt) -\int \wh
\vt_x^a\cE_{xa}d^nx)\rng - \label{z62'}\\
&& \qquad 
\lng\f^{a_1}_{x_1}\cdots \f^{a_k}_{x_k}\rng\lng {\rm Sp}(\wh \vt) -\int
\wh
\vt_x^a\cE_{xa}d^nx\rng =0.
\nonumber
\een
for complete Euclidean Green functions (\ref{z49}).

In particular, if $\vt$ is a variational supersymmetry of a Lagrangian
$L$ (e.g., the BRST supersymmetry of a gauge-fixing Lagrangian
\cite{jmp06}), the identities (\ref{z62'}) are the Ward identities 
\mar{bb11}\beq
\lng\wh \vt(\f^{a_1}_{x_1}\cdots \f^{a_k}_{x_k})\rng +
\lng\f^{a_1}_{x_1}\cdots \f^{a_k}_{x_k}{\rm Sp}(\wh \vt)\rng -  
\lng\f^{a_1}_{x_1}\cdots \f^{a_k}_{x_k}\rng\lng {\rm Sp}(\wh \vt)\rng =0, 
\label{bb11}
\eeq
generalizing the Ward
(Slavnov--Taylor) identities in the Yang--Mills gauge theory
\cite{slavn} (see Section 5 for an example of supersymmetric
Yang--Mills theory).

If $\vt=c^a\dr_a$, $c^a=$const, the identities (\ref{z62'}) take the form
\mar{bb13}\ben
&& \op\sum_{r=1}^k (-1)^{[a]([a_1]+\cdots+[a_{r-1}])}
\lng\f^{a_1}_{x_1}\cdots
\f^{a_{r-1}}_{x_{r-1}} \dl^{a_r}_a \f^{a_{r+1}}_{x_{r+1}}\cdots
\f^{a_k}_{x_k})\rng -  \label{bb13}\\
&& \qquad \lng\f^{a_1}_{x_1}\cdots \f^{a_k}_{x_k}(\int \cE_{xa}d^nx)\rng
+ 
\lng\f^{a_1}_{x_1}\cdots \f^{a_k}_{x_k}\rng\lng \int \wh
\cE_{xa}d^nx\rng =0. \nonumber
\een
One can think of them as being equations for complete Euclidean Green
functions, but they are not an Euclidean variant of he well-known
Schwinger--Dyson equations \cite{bog1}. For instance, they
identically hold if a Lagrangian
$L$ is quadratic. 

Clearly, the expressions (\ref{z62'}) -- (\ref{bb13}) are singular,
unless one follows regularization and renormalization procedures, which
however can induce additional anomaly terms.

\section{Lagrangian systems of even and odd fields}

As was mentioned above, we consider a Lagrangian
field system on $X=\Bbb R^n$, coordinated by $(x^\la)$.
Such a Lagrangian system is algebraically described in terms of the
following bigraded differential algebra (henceforth BGDA)
$\cP^*$ (\ref{bb1}) [1-3].

Let $Y\to X$ be an affine bundle coordinated by $(x^\la,y^i)$
whose sections are even classical fields, and let $Q\to X$ be a
vector bundle coordinated by $(x^\la,q^a)$ whose sections are odd
ones. Let $J^rY\to X$ and $J^rQ\to X$, $r=1,\ldots,$
be the corresponding $r$-order jet bundles, endowed with the
adapted coordinates $(x^\la, y^i_\La)$ and $(x^\la, q^a_\La)$,
respectively.
The index $r=0$
conventionally stands for $Y$ and $Q$. For each $r=0,\ldots,$ we
consider a graded manifold $(X,\cA_{J^rQ})$, whose body is $X$ and
the algebra of graded functions consists of sections of the exterior
bundle
\be
\w (J^rQ)^*=\Bbb R\op\oplus_X (J^rQ)^*\op\oplus_X\op\w^2
(J^rQ)^*\op\oplus_X\cdots,
\ee
where $(J^rQ)^*$ is the dual of a vector bundle $J^rQ\to X$. The
global basis for $(X,\cA_{J^rQ})$ is $\{x^\la,c^a_\La\}$,
$|\La|=0,\ldots,r$. Let us consider the graded commutative
$C^\infty(X)$-algebra $\cP^0$ generated by the even elements
$y^i_\La$ and the odd ones $c^a_\La$, $|\La|\geq 0$. The
collective symbols $s^a_\La$ stand for these elements together
with the symbol $[a]$ for their Grassmann parity.

Let $\gd \cP^0$ be the  Lie superalgebra of graded
derivations of the $\Bbb R$-algebra $\cP^0$, i.e.,
\be
u(ff')=u(f)f'+(-1)^{[u][f]}fu(f'), \qquad f,f'\in \cP^0,
\qquad u\in \gd\cP^0.
\ee
Its elements take the form
\mar{w5}\beq
 u=u^\la\dr_\la + \op\sum_{0\leq|\La|} u_\La^a\dr^\La_a,
 \qquad u^\la, u_\La^a
 \in \cP^0. \label{w5}
\eeq
With the Lie superalgebra $\gd \cP^0$, one can construct the
minimal Chevalley--Eilenberg differential calculus
\mar{bb1}\beq
0\to \Bbb R\to \cP^0 \ar^d \cP^1\ar^d\cdots
\cP^2\ar^d\cdots \label{bb1}
\eeq
over the $\Bbb R$-algebra $\cP^0$. It is the above mentioned  BGDA $\cP^*$
with the basis $\{s^a\}$. Its elements $\si\in \cP^k$ are graded
$\cP^0$-linear
$k$-forms 
\be
 \si= \op\sum \si_{a_1\ldots a_r\la_{r+1}\ldots\la_k}^{\La_1\ldots
\La_r} ds_{\La_1}^{a_1}\w\cdots\w ds_{\La_r}^{a_r}\w
dx^{\la_{r+1}}\w\cdots \w dx^{\la_k}
\ee
on $\gd \cP^0$ with values in $\cP^0$. The
graded exterior product $\w$ and the graded exterior differential,
obey the relations
\be
 \si\w\si' =(-1)^{|\si||\si'| +[\si][\si']}\si'\w
\si, \qquad  d(\si\w\si')= d\si\w\si' +(-1)^{|\si|}\si\w d\si',
\ee
where $|.|$ denotes the form degree. 
By $\cO^*X$ is denoted the graded differential algebra of exterior forms
on $X$. There is the natural monomorphism $\cO^*X\to \cP^*$.

Given a graded derivation $u$ (\ref{w5}) of the $\Bbb R$-algebra
$\cP^0$, the interior product $u\rfloor\si$ and the Lie
derivative $\bL_u\si$, $\si\in\cP^*$, obey the relations
\be
&& u\rfloor(\si\w\si')=(u\rfloor \si)\w\si'
+(-1)^{|\si|+[\si][u]}\si\w(u\rfloor\si'), \qquad \si,\si'\in
\cP^*, \\
&& \bL_u\si=u\rfloor d\si+ d(u\rfloor\si), \qquad
\bL_u(\si\w\si')=\bL_u(\si)\w\si' +(-1)^{[u][\si]}\si\w\bL_u(\si').
\ee
For instance, let us denote $d_\la\si=\bL_{d_\la}\si$ 

The BGDA $\cP^*$ is
decomposed into $\cP^0$-modules $\cP^{k,r}$ of
$k$-contact and $r$-horizontal graded forms
\be
\si=\op\sum_{0\leq|\La_i|}\si^{\La_1\ldots \La_k}_{a_1\ldots a_k
\m_1\ldots\m_r} \th^{a_1}_{\La_1}\w\cdots\w\th^{a_k}_{\La_k}\w
dx^{\m_1}\w\cdots\w
dx^{\m_r}, \qquad \th^a_\La=ds^a_\La -s^a_{\la+\La}dx^\la.
\ee
Accordingly, the graded exterior differential on $\cP^*$
falls into the sum $d=d_V+d_H$ of the vertical and total
differentials where $d_H\si= dx^\la\w d_\la\si$.
The differentials
$d_H$ and $d_V$ and the graded variational operator $\dl$ split
the BGDA $\cP^*$ into the graded variational bicomplex
\cite{cmp04,jmp05,barn}. Its elements $L\in \cP^{0,n}$ and $\dl L
\in\cP^{1,n}$ (\ref{0709}) are a Lagrangian and its Euler--Lagrange
operator, respectively.

A graded derivation $u$ (\ref{w5}) is called contact if the Lie
derivative $\bL_u$ preserves the ideal of contact graded forms of
the BGDA $\cP^*$. Here, we restrict our consideration to
vertical contact graded derivations, vanishing on $\cO^*X$. Such a
derivation takes the form (\ref{0672}), and 
is determined by its first summand $\up=\up^a\dr_a$. The Lie
derivative
$\bL_\vt L$ of a Lagrangian $L$ along a vertical contact graded
derivation $\vt$ (\ref{0672}) admits the decomposition (\ref{g107})
\mar{g107'}\beq
\bL_\vt L= \up\rfloor\dl L +d_H\si. \label{g107'}
\eeq
One says that an odd vertical
contact graded derivation
$\vt$ (\ref{0672}) is a variational supersymmetry of a Lagrangian $L$ if
the Lie derivative $\bL_\vt L$ is $d_H$-exact.

\section{Quantization}

Hereafter, let $(\cP^*,L)$ be either an original nondegenerate
Lagrangian system or the nondegenerate Lagrangian system derived from an
original one by means of the above mentioned BV
procedure. Let us quantize this nondegenerate Lagrangian system
$(\cP^*,L)$. Though our results lie in the framework of
perturbative QFT, we start with algebraic QFT.

In algebraic QFT, a quantum field system is characterized by a
topological $^*$-algebra $A$ and a continuous positive form $f$ on
$A$ \cite{borch,hor}.  For the sake of simplicity, let us
consider even scalar fields on the Minkowski space $X=\Bbb R^n$. One
associates to them the Borchers algebra $A_\Phi$ of tensor products of the
nuclear Schwartz space $\Phi=S(\Bbb R^n)$ of smooth complex
functions of rapid decreasing at infinity on $\Bbb R^n$. The
topological dual of $S(\Bbb R^n)$ is the space $S'(\Bbb R^n)$ of
tempered distributions \cite{piet,bog}.

Let $\Bbb R_n$ denote the dual of $\Bbb R^n$ coordinated by
$(p_\la)$. The Fourier transform
\mar{spr460,1}\ben
&& \f^F(p)=\int \f(x)e^{ipx}d^nx, \qquad px=p_\la x^\la,
\label{spr460}\\
&& \f(x)=\int \f^F(p)e^{-ipx}d_np, \qquad d_np=(2\pi)^{-n}d^np,
\label{spr461}
\een
yields an isomorphism between the spaces $S(\Bbb R^n)$ and $S(\Bbb
R_n)$. The Fourier transform of distributions is defined by the
condition
\be
\int \psi(x)\f(x)d^nx=\int \psi^F(p)\f^F(-p)d_np,
\ee
and is written in the form (\ref{spr460}) -- (\ref{spr461}). It
provides an isomorphism between the spaces of distributions
$S'(\Bbb R^n)$ and $S'(\Bbb R_n)$.

Since $\op\ot^kS(\Bbb R^n)$ is dense in $S(\Bbb R^{nk})$, a state
$f$ of the Borchers algebra $A_\Phi$ is represented by
distributions
\be
f_k(\phi_1\cdots\phi_k)=\int
W_k(x_1,\ldots,x_k)\phi_1(x_1)\cdots\phi_k(x_k) d^nx_1\ldots
d^nx_k, \qquad W_k\in S'(\Bbb R^{nk}).
\ee
In particular, the $k$-point Wightman functions $W_k$ describe
free fields in the Minkowski space. The complete Green functions
characterize quantum fields created at some instant and
annihilated at another one. They are given by the chronological
functionals
\mar{1260}\ben
&& f^c(\phi_1\cdots\phi_k)=\int
W^c_k(x_1,\ldots,x_k)\phi_1(x_1)\cdots\phi_k(x_k)
d^nx_1\ldots d^nx_k, \label{1260}\\
&& W^c_k(x_1,\ldots,x_k)= \op\sum_{(i_1\ldots
i_k)}\th(x^0_{i_1}-x^0_{i_2})
\cdots\th(x^0_{i_{k-1}}-x^0_{i_k})W_k(x_1,\ldots,x_k), \quad
W_k\in S'(\Bbb R^{nk}),\nonumber
\een
where $\th$ is the step function, and the sum runs through all
permutations $(i_1\ldots i_k)$ of the numbers $1,\ldots,k$.
However, the chronological functionals (\ref{1260}) need not be
continuous and positive. At the same time, they issue from the
Wick rotation of Euclidean states of the Borchers algebra $A_\Phi$
describing quantum fields in an interaction zone
\cite{sard91,sard02}. Since the chronological functionals
(\ref{1260}) are symmetric, these Euclidean states are states of the
corresponding commutative tensor algebra $B_\Phi$. This is the enveloping
algebra of the Lie algebra of the group $T(\Phi)$ of translations in
$\Phi$. Therefore one can obtain a state of $B_\Phi$ as a vector form of a
strong-continuous unitary cyclic representation of
$T(\Phi)$ \cite{gel}. Such a representation is characterized by a
positive-definite continuous generating function $Z$ on $\Phi$. By
virtue of the Bochner theorem \cite{gel}, this function is the
Fourier transform
\mar{031}\beq
Z(\phi)=\op\int_{\Phi'}\exp[i \langle\phi,w\rangle]d\mu(w)
\label{031}
\eeq
of a positive measure $\mu$ of total mass 1 on the topological
dual $\Phi'$ of $\Phi$. If the function $\alpha\to Z(\alpha\phi)$
on $\Bbb R$ is analytic at 0 for each $\phi\in \Phi$, a state $F$
of $B_\Phi$ is given by the expression
\mar{w0}\beq
F_k(\phi_1\cdots\phi_k)=i^{-k}\frac{\dr}{\dr \al^1}
\cdots\frac{\dr}{\dr\alpha^k}Z(\alpha^i\phi_i)|_{\alpha^i=0}=\int\langle
\phi_1,w\rangle\cdots\langle \phi_k,w \rangle d\mu(w). \label{w0}
\eeq
Then one can regard $Z$ (\ref{031}) as a generating functional of complete
Euclidean Green functions $F_k$ (\ref{w0}). However, it is a
problem is that, if a field Lagrangian is a polynomial of degree
exceeding two, a generating functional $Z$ (\ref{031}) and Green
functions fail to be written in
an explicit form.

Therefore, let us quantize the above mentioned nondegenerate
Lagrangian system $(\cP^*,L)$ in the framework of
perturbative QFT. We assume that $L$ is a Lagrangian of
Euclidean fields on $X=\Bbb R^n$. The key point is that an algebra
of Euclidean quantum fields is graded commutative, and there are
homomorphisms of the graded commutative algebra $\cP^0$ of
classical fields to this algebra.

Let $\cQ$ be the graded complex vector space whose basis is the
basis $\{s^a\}$ for the BGDA $\cP^*$.  Let us consider
the tensor product
\mar{z40}\beq
\Phi=\cQ\ot S'(\Bbb R^n) \label{z40}
\eeq
of the graded vector space $\cQ$ and the space $S'(\Bbb R^n)$ of
distributions on $\Bbb R^n$. One can think of elements $\Phi$
(\ref{z40}) as being $\cQ$-valued distributions on $\Bbb R^n$. Let
$T(\Bbb R^n)\subset S'(\Bbb R^n)$ be a subspace of
functions $\exp\{ipx'\}$, $p\in \Bbb R_n$, which are generalized
eigenvectors of translations in $\Bbb R^n$ acting on $S(\Bbb
R^n)$. We denote $\f^a_p=s^a\ot\exp\{ipx'\}$. Then any element
$\f$ of $\Phi$ can be written in the form
\mar{z42}\beq
\f(x')=s^a\ot\f_a(x')=\int \f_a(p)\f_p^a d_np, \label{z42}
\eeq
where $\f_a(p)\in S'(\Bbb R_n)$ are the Fourier transforms of
$\f_a(-x')$. For instance, there are the $\cQ$-valued
distributions
\mar{z32,43}\ben
&& \f^a_x(x')=\int \f^a_p e^{-ipx}d_np=s^a\ot \dl(x-x'),
\label{z32}\\
&& \f^a_{x\La}(x')=\int (-i)^k p_{\la_1}\cdots p_{\la_k}\f^a_p
e^{-ipx}d_np. \label{z43}
\een

In the framework of perturbative Euclidean QFT, we associate to a
nondegenerate Lagrangian system $(\cP^*,L)$ the graded
commutative tensor algebra $B_\Phi$ generated by elements of the
graded vector space $\Phi$ (\ref{z40}) and the following state
$\lng.\rng$ of $B_\Phi$. 

For
any $x\in X$, there is a homomorphism $\g_x$ (\ref{z45})
of the algebra $\cP^0$ of classical fields to the algebra $B_\Phi$
which sends the basis elements $s^a_\La\in \cP^0$ to the
elements $\f^a_{x\La}\in B_\Phi$, and replaces
coefficient functions $f$ of elements of $\cP^0$ with their
values $f(x)$ at a point $x$. Then the above mentioned state $\lng.\rng$
of $B_\Phi$ is given by symbolic functional integrals
\mar{z31,',47}\ben
&& \lng \f_1\cdots \f_k\rng=\frac{1}{\cN}\int \f_1\cdots \f_k
\exp\{-\int \cL(\f^a_p)d^nx\}\op\prod_p
[d\phi_p^a], \label{z31}\\
&& \cN=\int \exp\{-\int \cL(\f^a_p)d^nx\}\op\prod_p
[d\phi_p^a], \label{z31'}\\
&&
\cL(\f^a_p)=\cL(\f^a_{x\La})=\cL(x,\g_x(s^a_\La)),
\label{z47}
\een
where $\f_i$ and $\g_x(s^a_\La)=\f^a_{x\La}$ are given by the
formulas (\ref{z42}) and (\ref{z43}), respectively. Clearly, the
expression
$\cL(\f^a_p)$ (\ref{z47}) is not local.  The forms (\ref{z31}) are
expressed both into the forms
\mar{z48}\beq
 \lng\f^{a_1}_{p_1}\cdots \f^{a_k}_{p_k}\rng=\frac{1}{\cN}\int
\f^{a_1}_{p_1}\cdots \f^{a_k}_{p_k} \exp\{-\int
\cL(\f^a_p)d^nx\}\op\prod_p [d\phi_p^a], \label{z48}
\eeq
and the forms $\lng\f^{a_1}_{x_1}\cdots \f^{a_k}_{x_k}\rng$ (\ref{z49})
which provide Euclidean Green functions. It should be emphasized that, in
contrast with a measure $\m$ in the expression (\ref{031}), the term
$\op\prod_p [d\phi_p^a]$ in the formulas (\ref{z31}) -- (\ref{z31'}) fail
to be a true measure on $T(\Bbb R^n)$ because the Lebesgue measure on
infinite-dimensional vector spaces need not exist. Nevertheless, treated
as generalization of Berezin's finite-dimensional integrals \cite{ber},
the functional integrals (\ref{z49}) and (\ref{z48})  restart Euclidean
Green functions in the Feynman diagram technique. Certainly, these Green
functions are singular, unless regularization and renormalization
techniques are involved.

\section{The Green function identities}

Since a graded derivation $\vt$ (\ref{0672}) of the algebra $\cP^0$ is a
$C^\infty(X)$-linear morphism over
$\id X$, it induces the graded derivation
\mar{z50}\beq
\wh \vt_x= \g_x\circ \wh \vt\circ \g^{-1}_x: \f^a_{x\La}\to (x,
q^a_\La))\to \wh \vt^a_\La(x,q^b_\Si)\to \wh
\vt^a_\La(x,\g_x(q^b_\Si))=\wh \vt^a_{x\La}(\f^b_{x\Si}) \label{z50}
\eeq
of the range $\g_x(\cP^0)\subset B_\Phi$ of the homomorphism
$\g_x$ (\ref{z45}) for each $x\in \Bbb R^n$.
The maps $\wh \vt_x$
(\ref{z50}) yield the maps
\be
&& \wh \vt_p: \f^a_p=\int \f^a_x
e^{ipx}d^nx \to \int \wh \vt_x(\f^a_x)e^{ipx}d^nx= \int \wh
\vt^a_x(\f^b_{x\Si})e^{ipx}d^nx = \\
&& \qquad  \int \wh \vt_x^a(\int (-i)^kp'_{\si_1}\cdots
p'_{\si_k}\f^b_{p'}e^{-ip'x}d_np') e^{ipx}d^nx= \wh \vt^a_p, \qquad
p\in\Bbb R_n,
\ee
and, as a consequence, the graded derivation
\be
\wh \vt(\f)=\int\f_a(p)\wh \vt(\f^a_p) d_np= \int\f_a(p)\wh \vt^a_pd_np
\ee
of the algebra $B_\Phi$. It can be written in the symbolic form
\mar{z53,4}\ben
&& \wh \vt= \int u^a_p\frac{\dr}{\dr \f^a_p}d_np, \qquad \frac{\dr
\f^b_{p'}}{\dr \f^a_p} =\dl^b_a\dl(p'-p), \label{z53}\\
&& \wh \vt= \int u^a_x\frac{\dr}{\dr \f^a_x}d^nx, \qquad \frac{\dr
\f^b_{x'\La}}{\dr \f^a_x}= \dl^b_a\frac{\dr}{\dr x'^{\la_1}}
\cdots \frac{\dr}{\dr x'^{\la_k}} \dl(x'-x). \label{z54}
\een

Let $\al$ be an odd element. Then $\wh U$ (\ref{bb20}) is an 
automorphism  of the algebra $B_\Phi$, and can provide a change of
variables depending on $\al$ as a parameter in the functional integrals
(\ref{z49}) and (\ref{z48})
\cite{ber}. This automorphism yields a new state
$\lng.\rng'$ of $B_\Phi$ given by the relations (\ref{bb10}) and
\be
&& \lng \f_1\cdots \f_k\rng= \lng \wh U(\f_1)\cdots \wh
U(\f_k)\rng'= \\
&& \qquad \frac{1}{\cN'}\int \wh U(\f_1)\cdots \wh U(\f_k)
\exp\{-\int \cL_{GF}(\wh U(\f^a_p))d^nx\}\op\prod_p [d\wh
U(\phi_p^a)], \\
&& \cN'=\int \exp\{-\int \cL_{GF}(\wh U(\f^a_p))d^nx\}\op\prod_p
[d\wh U(\phi_p^a)].
\ee
Let us apply these relations to the Green functions
(\ref{z49}) and (\ref{z48}).

Using the first variational formula (\ref{g107'}), the equalities
(\ref{bb21}) and 
\be
\op\prod_x[d\wh U(\phi_x^a)]=(1+\al\int \frac{\dr \wh
\vt^a_x}{\dr \f^a_x}d^nx)\op\prod_x[d\phi_x^a]=(1+\al {\rm Sp}(\wh
\vt)) \op\prod_x[d\phi_x^a],
\ee
one comes to the desired identities (\ref{z62'}) and the similar
identities for the Green functions $\lng\f^{a_1}_{p_1}\cdots
\f^{a_k}_{p_k}\rng$. If $\vt$ is a variational supersymmetry of $L$, we
obtain the above mentioned Ward identities 
\mar{z61,2}\ben
&& \lng\wh \vt(\f^{a_1}_{p_1}\cdots \f^{a_k}_{p_k})\rng
+\lng\f^{a_1}_{p_1}\cdots \f^{a_k}_{p_k}{\rm Sp}(\wh \vt)\rng- \lng
\f^{a_1}_{p_1}\cdots \f^{a_k}_{p_k}\rng\lng{\rm Sp}(\wh \vt)\rng =0,
\label{z61}\\
&& \op\sum_{i=1}^k(-1)^{[a_1]+\cdots +[a_{i-1}]}
\lng\f^{a_1}_{p_1}\cdots \f^{a_{i-1}}_{p_{i-1}}\wh \vt^{a_i}_{p_i}
\f^{a_{i+1}}_{p_{i+1}}\cdots \f^{a_k}_{p_k}\rng +\nonumber\\
&& \qquad \lng\f^{a_1}_{p_1}\cdots \f^{a_k}_{p_k}\int \frac{\dr
\wh \vt^a_p}{\dr \f^a_p}d_np\rng - \lng\f^{a_1}_{p_1}\cdots
\f^{a_k}_{p_k}\rng\lng\int \frac{\dr \wh \vt^a_p}{\dr
\f^a_p}d_np\rng=0,
\nonumber\\
&& \lng\wh \vt(\f^{a_1}_{x_1}\cdots \f^{a_k}_{x_k})\rng
+\lng\f^{a_1}_{x_1}\cdots \f^{a_k}_{x_k}{\rm Sp}(\wh \vt)\rng- \lng
\f^{a_1}_{x_1}\cdots \f^{a_k}_{x_k}\rng\lng{\rm Sp}(\wh \vt)\rng =0,
\label{z62}\\
&& \op\sum_{i=1}^k(-1)^{[a_1]+\cdots +[a_{i-1}]}
\lng\f^{a_1}_{x_1}\cdots \f^{a_{i-1}}_{x_{i-1}}\wh \vt^{a_i}_{x_i}
\f^{a_{i+1}}_{x_{i+1}}\cdots \f^{a_k}_{x_k}\rng +\nonumber\\
&&\qquad \lng\f^{a_1}_{x_1}\cdots \f^{a_k}_{x_k}\int \frac{\dr \wh
\vt^a_x}{\dr \f^a_x}d^nx\rng - \lng\f^{a_1}_{x_1}\cdots
\f^{a_k}_{x_k}\rng\lng\int \frac{\dr \wh \vt^a_x}{\dr
\f^a_x}d^nx\rng=0. \nonumber
\een

A glance at the expressions (\ref{z61}) -- (\ref{z62}) shows that
these Ward identities generally contain anomaly because the
measure terms of symbolic functional integrals need not be 
$\wh\vt$-invariant. If Sp$(\wh \vt)$ is either a finite or infinite
number, the Ward identities
\mar{z63,4}\ben
&& \lng\wh \vt(\f^{a_1}_{p_1}\cdots \f^{a_k}_{p_k})\rng =
\op\sum_{i=1}^k(-1)^{[a_1]+\cdots +[a_{i-1}]}
\lng\f^{a_1}_{p_1}\cdots \f^{a_{i-1}}_{p_{i-1}}\wh \vt^{a_i}_{p_i}
\f^{a_{i+1}}_{p_{i+1}}\cdots \f^{a_k}_{p_k}\rng=0, \label{z63}\\
&& \lng\wh \vt(\f^{a_1}_{x_1}\cdots \f^{a_k}_{x_k})\rng=
\op\sum_{i=1}^k(-1)^{[a_1]+\cdots +[a_{i-1}]}
\lng\f^{a_1}_{x_1}\cdots \f^{a_{i-1}}_{x_{i-1}}\wh \vt^{a_i}_{x_i}
\f^{a_{i+1}}_{x_{i+1}}\cdots \f^{a_k}_{x_k}\rng =0 \label{z64}
\een
are free of this anomaly.

\section{Supersymmetric Yang--Mills theory}

Let $\cG=\cG_0\oplus \cG_1$ be a finite-dimensional real Lie
superalgebra with a basis $\{e_r\}$, $r=1,\ldots,m,$ and real
structure constants $c^r_{ij}$. Further, the
Grassmann parity of $e_r$ is denoted by $[r]$. Recall the standard
relations
\be
&& c^r_{ij}=-(-1)^{[i][j]}c^r_{ji}, \qquad [r]=[i]+[j],\\
&& (-1)^{[i][b]}c^r_{ij}c^j_{ab} + (-1)^{[a][i]}c^r_{aj}c^j_{bi} +
(-1)^{[b][a]}c^r_{bj}c^j_{ia}=0.
\ee
Let us also introduce the modified structure constants
\be
\ol c^r_{ij}=(-1)^{[i]}c^r_{ij}, \qquad \ol
c^r_{ij}=(-1)^{([i]+1)([j]+1)}\ol c^r_{ji}. 
\ee
Given the universal enveloping algebra $\ol \cG$ of $\cG$, we
assume that there is an invariant even quadratic element
$h^{ij}e_ie_j$ of $\ol\cG$ such that the matrix $h^{ij}$ is
nondegenerate. 

The Yang--Mills theory of gauge potentials on $X=\Bbb R^n$
associated to the Lie superalgebra $\cG$ is described by the BGDA
$\cP^*$ where
\be
Q=(X\times \cG_1)\op\ot_X T^*X, \qquad Y= (X\times \cG_0)\op\ot_X
T^*X.
\ee
Its basis is $\{a^r_\la\}$, $[a^r_\la]=[r]$. There is the
canonical decomposition of the first jets of its elements
\be
a^r_{\la\m}=\frac12(\cF^r_{\la\m} +
\cS^r_{\la\m})=\frac12(a^r_{\la\m}-a^r_{\m\la} +c^r_{ij}a^i_\la
a^j_\m) +\frac12(a^r_{\la\m}+ a^r_{\m\la} -c^r_{ij}a^i_\la
a^j_\m).
\ee
Then the Euclidean Yang--Mills Lagrangian takes the form
\be
L_{YM}=\frac14
h_{ij}\eta^{\la\m}\eta^{\bt\nu}\cF^i_{\la\bt}\cF^j_{\m\nu}d^nx,
\ee
where $\eta$ is the Euclidean metric on $\Bbb R^n$. It degenerate
because its variational derivatives $\cE_r^\la$ obey the irreducible
Noether identities
\be
 -c^r_{ji}a^i_\la\cE_r^\la - d_\la\cE_j^\la=0.
\ee
Therefore, the above mentioned BV
procedure must be applied to this field model \cite{jmp06}. As a result,
the original BGDA
$\cP^*$ is enlarged to the BGDA
$\ol P^*$ whose basis
\be
\{a^r_\la, c^r, c^*_r\}, \qquad [c^r]=([r]+1){\rm
mod}\,2, \qquad [c^*_r]=[c^r],
\ee
consists of gauge potentials $a^r_\la$, ghosts $c^r$ and antighosts
$c^*_r$. The
final gauge-fixing Lagrangian reads
\be
&& L_{GF}=L_{YM} +c^*_r M_j^r c^jd^nx +\frac18
h_{ij}\eta^{\la\m}\eta^{\bt\nu}\cS^i_{\la\m}\cS^j_{\bt\nu}d^nx=\\
&& \qquad L_{YM} + [(-1)^{[r]+1}\eta^{\la\m}c^*_r
d_\m(-c^r_{ij}c^ja^i_\la + c^r_\la) + \frac12
h_{ij}\eta^{\la\m}\eta^{\bt\nu}a^i_{\la\m}a^j_{\bt\nu}]d^nx,
\ee
where
\be
M_j^r=(-1)^{[r]+1}\eta^{\la\m}(-(-1)^{[i]([j]+1}c^r_{ij}(a^i_{\m\la}+
a^i_\la d_\m) + \dl^r_j d_{\m\la})
\ee
is a second order differential operator acting on the ghosts
$c^j$. The Lagrangian  $L_{GF}$ possesses the variational BRST symmetry
\be
\vt= (-c^r_{ij}c^ja^i_\la + c^r_\la)\frac{\dr}{\dr a_\la^r}
-\frac12 \ol c^r_{ij}c^ic^j\frac{\dr}{\dr c^r} +(-1)^{[j]}
h_{ij}\eta^{\la\m}a^i_{\la\m}\frac{\dr}{\dr c^*_j}.
\ee

Quantizing this Lagrangian system in the framework of Euclidean
perturbative QFT, we come to the graded commutative tensor algebra
$B_\Phi$ generated by the elements $\{a_{x\la}^r, c_x^r,
c^*_{xr}\}$. Its state $\lng .\rng$ is given by
functional integrals
\be
&& \lng \f\rng=\frac{1}{\cN}\int \f \exp\{-\int
\cL_{GF}(a^r_{x\La\la}, c^r_{x\La}, c^*_{xr})d^nx\}\op\prod_x
[da^r_{x\la}][dc^r_x][dc^*_{xr}], \\
&& \cN=\int \exp\{-\int \cL_{GF}(a^r_{x\La\la}, c^r_{x\La},
c^*_{xr})d^nx\}\op\prod_x [da^r_{x\la}][dc^r_x][dc^*_{xr}],\\
&& \cL_{GF}=\cL_{YM} + (-1)^{[r]+1}\eta^{\la\m}c^*_{xr}
d_\m(-c^r_{xij}c^j_xa^i_{x\la} + c^r_{x\la}) + \frac12
h_{ij}\eta^{\la\m}\eta^{\bt\nu}a^i_{x\la\m}a^j_{x\bt\nu}.
\ee
Accordingly, the quantum BRST transformation (\ref{z54}) reads
\be
\wh \vt= \int[(-c^r_{ij}c^j_xa^i_{x\la} + c^r_{x\la})\frac{\dr}{\dr
a_{x\la}^r} -\frac12 \ol c^r_{ij}c^i_xc^j_x\frac{\dr}{\dr c^r_x}
+(-1)^{[j]} h_{ij}\eta^{\la\m}a^i_{x\la\m}\frac{\dr}{\dr
c^*_{xj}}]d^nx.
\ee
It is readily observed that Sp$(\wh \vt)=0$. Therefore, we obtain
the Ward identities $\lng\wh \vt(\f)\rng=0$ (\ref{z64}) without
anomaly.

\end{document}